\documentclass{ws-procs9x6}

\usepackage{epsfig}
\usepackage{graphics}
\usepackage{amsfonts}
\usepackage{amsbsy}
\usepackage{amssymb}
\usepackage[mathscr]{eucal}

\DeclareRobustCommand*\cal{\relax\mathcal}

\def\be{\begin{equation}}
\def\ee{\end{equation}}

\begin{document}


\title{Baryon Resonance Phenomenology}

\author{ \vspace{-1em}
I.C. Cloet,        
D.B. Leinweber and 
A.W. Thomas        
}

\address{Special Research Centre for the Subatomic Structure of 
         Matter and      \\
  Department of Physics and Mathematical Physics, University of Adelaide,
  SA 5005, Australia} 

\maketitle

\vspace{-18em} \hfill ADP-02-98/T536  \\
\vspace{17em}
\vspace{-17em} \hfill  nucl-th/0211027 \\
\vspace{13em}

\abstracts{The Japan Hadron Facility will provide an unprecedented
opportunity for the study of baryon resonance properties.  This talk
will focus on the chiral nonanalytic behaviour of magnetic moments
exclusive to baryons with open decay channels.  To illustrate the
novel features associated with an open decay channel, we consider the
``Access'' quark model, where an analytic continuation of chiral
perturbation theory is employed to connect results obtained using the
constituent quark model in the limit of SU(3)-flavour symmetry to
empirical determinations.}


\vspace{-2em}

\section{Introduction}

The Japan Hadron Facility will present new opportunities for the
investigation of baryon resonance properties.  In particular, access
to the hyperons of the baryon decuplet will be unprecedented.  This
talk serves to highlight the novel and important aspects of QCD that
can be explored through an experimental program focusing on
decuplet-baryon resonance phenomenology.

To highlight the new opportunities, it is sufficient to address the
magnetic moments of the charged $\Delta$ baryons of the decuplet.  The
magnetic moments of these baryons have already caught the attention of
experimentalists and hold the promise of being accurately measured in
the foreseeable future.  Experimental estimates exist for the
${\Delta^{++}}$ magnetic moment, based on the reaction $\pi^+\ p\ \to
\pi^+\ \gamma'\ p$.  The Particle Data Group\cite{PDT} provides the
range 3.7--7.5~$\mu_N$ for the ${\Delta^{++}}$ magnetic moment with
the two most recent experimental results of $4.52 \pm 0.50 \pm
0.45\ \mu_N$\cite{Bosshard} and $6.14 \pm 0.51\ \mu_N$\cite{LopezCastro:2000cv} .  
In principle, the $\Delta^+$ magnetic moment can be
obtained from the reaction $\gamma\ p\ \to \pi^o\ \gamma'\ p$, as
demonstrated at the Mainz microtron.\cite{Kotulla}  An experimental
value for the $\Delta^+$ magnetic moment appears imminent.

Recent extrapolations of octet baryon magnetic
moments\cite{CLT,LLT,HLT} have utilized an analytic continuation of
the leading nonanalytic (LNA) structure of Chiral Perturbation Theory
($\chi$PT), as the extrapolation function.  The unique feature of this
extrapolation function is that it contains the correct chiral
behaviour as $m_q \to 0$ while also possessing the Dirac moment mass
dependence in the heavy quark mass regime.

The extrapolation function utilized here has these same features,
however we move beyond the previous approach by incorporating not only
the LNA but also the next-to-leading nonanalytic (NLNA) structure of
$\chi$PT in the extrapolation function.  Incorporating the NLNA terms
contributes little to the octet baryon magnetic moments, however it
proves vital for decuplet baryons.  The NLNA terms contain information
regarding the branch point at $m_{\pi} = M_\Delta - M_N$, associated
with the $\Delta \to N \pi$ decay channel and play a significant role
in decuplet-baryon magnetic moments.

\section{Leading and Next-to-Leading Nonanalytic Behavior}
\label{sec:chiPT}

We begin with the chiral expansion for decuplet baryon magnetic
moments.\cite{BM}  The LNA and NLNA behaviour is given by
\begin{equation}
q_i\, {\cal G}_K + \frac{2}{3} I_3^i({\cal G}_{\pi}-{\cal G}_K)\, ,
\label{CPT}
\end{equation}
where $q_i$ and $I_3^i$ are charge and isospin respectively, and 
${\cal G}_j~(j={\pi},K)$ is given by
\begin{equation}
{\cal G}_j = \frac{-M_N}{32{\pi} f_j{}^2} \left\{\frac{4}{9}{\cal H}^2
              {\cal F}(0,m_j,\mu_j) +  
              {\cal C}^2{\cal F}(-{\delta}_N,m_j,\mu_j)\right\}\, .
\label{GJ}
\end{equation}
$\cal H$ describes the meson coupling to decuplet baryons and $\cal C$
describes octet-decuplet transitions.  We take ${\cal H} = -2.2$ and
${\cal C} = -1.2$.  Here we omit the Roper\cite{BM} as this transition
requires significant excitation energy and is strongly suppressed by
the finite size of the meson source.  The octet-decuplet mass
splitting $\delta_N$ is assigned its average value
\begin{displaymath}
\delta_N=M_{10}-M_{8}=1377-1151=+226~{\rm MeV}.
\end{displaymath}  
We take $f_{\pi}=93~\rm MeV$ and $f_K = 112~\rm MeV$. The function
${\cal F}(\delta,m,\mu)$ has the form\footnote {This definition for
${\cal F}(\delta,m,\mu)$ corrects a sign error in Ref.~\cite{BM}.  It
differs by an overall minus sign and suppresses additive constants
which are irrelevant in our analysis.  As $\delta \to 0$, ${\cal
F}(\delta,m_\pi,\mu) \to m_\pi$.}
\begin{eqnarray}
m &>& \mid{\delta}\mid\, , \nonumber \\
&&{\cal F}(\delta,m,\mu) = -\frac{\delta}{\pi}\ln 
\left(\frac{m^2}{\mu^2}\right) + \frac{2}{\pi} \sqrt{m^2-{\delta}^2}
\left(\frac{\pi}{2} - \tan^{-1}
\frac{\delta}{\sqrt{m^2-{\delta}^2}}\right), 
\nonumber
\\
m &<& \mid{\delta}\mid\, , \nonumber \\ 
&&{\cal F}(\delta,m,\mu) = -\frac{\delta}{\pi}\ln
\left(\frac{m^2}{\mu^2}\right) +\frac{1}{\pi} \sqrt{{\delta}^2 - m^2} 
\ \ln\left(\frac{\delta-\sqrt{\delta^2-m^2}}{\delta+\sqrt{\delta^2-m^2}}\right)
\, .
\label{F}
\end{eqnarray}  
Hence the LNA and NLNA behaviour of decuplet magnetic moments is given by
\begin{equation}
\chi_{\pi}\ m_{\pi} + \chi_{K}\ m_{K} + \chi_{\pi}'\ {\cal F}(-{\delta}_N,m_{\pi},{\mu}_{\pi})
                                                  + \chi_{K}'\ {\cal F}(-{\delta}_N,m_{K},{\mu}_{K})\, ,
\label{PT1}
\end{equation}
where 
\vspace{0.2em}
\begin{eqnarray}
\chi_{\pi} = \frac{M_N\ {\cal H}^2}{{\pi}\ (f_{\pi})^2}
\left(\frac{-I_3^i}{108}\right)\, , \hspace{9.5mm} \qquad
\chi_{\pi}'= \frac{M_N\ {\cal C}^2}{{\pi}\ (f_{\pi})^2}
\left(\frac{-I_3^i}{48}\right)\, , \hspace{9.0mm}\nonumber \\   
\chi_{K}   = \frac{M_N\ {\cal H}^2}{{\pi}\ (f_K)^2}
\left(\frac{I^i_3}{108}-\frac{q_i}{72}\right)\, ,\qquad
\chi_{K}'  = \frac{M_N\ {\cal C}^2}{{\pi}\ (f_K)^2}
\left(\frac{I^i_3}{48} -
\frac{q_i}{32}\right)\hspace{1.5mm}\, .\label{CC} 
\end{eqnarray}
The values for the above chiral coefficients, Eqs.~(\ref {CC}),
describing the strength of various meson dressings of the $\Delta$
baryons, are summarized in Table~\ref{table:DCC} for the four $\Delta$
baryons of the decuplet.

\begin{table}[b]
\tbl{The baryon chiral coefficients for the four $\Delta$ baryons of
the decuplet.  Coefficients are calculated with ${\cal H}=-2.2$ and
${\cal C}=-1.2$, where ${\cal H}$ is the decuplet-decuplet coupling
constant and ${\cal C}$ is the decuplet-octet coupling constant.  We
have suppressed the kaon loop contribution by using
$f_{K}=1.2~f_{\pi}$ and $f_{\pi} = 93$ MeV.} 
{\footnotesize 
\tabcolsep20pt
\begin{tabular}{|l|c|c|c|c|}
\hline
{} &{} &{} &{} &{} \\[-1.5ex]
               &$\Delta^{++}$  &$\Delta^{+}$  &$\Delta^{0}$  &$\Delta^{-}$ \\
\hline
{} &{} &{} &{} &{} \\[-1.5ex]
$\chi_{\pi}$   &$-$2.33        &$-$0.777      &$+$0.777      &$+$2.33 \\
$\chi_K$       &$-$1.61        &$-$1.070      &$-$0.535      &0 \\
$\chi_{\pi}'$  &$-$1.56        &$-$0.518      &$+$0.518      &$+$1.56 \\
$\chi_K'$      &$-$1.08        &$-$0.719      &$-$0.361      &0 \\[+1.5ex]
\hline
\end{tabular}
}
\label{table:DCC}
\end{table}

\section{Analytic Continuation of $\chi$PT}
\vspace{0.1em}

\begin{figure}[t]
\begin{center}
{\epsfig{file=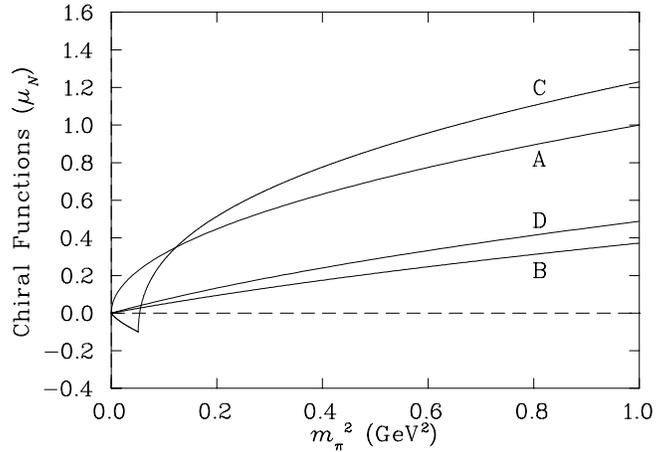, height=8.5cm, angle=90}}
\caption{Plots of the four chiral expansion functions (without the
chiral coefficient pre-factors) of Eq.~(\ref{PT2}), labeled A, B, C, D
in Eq.~(\ref{PT2}).
\vspace{-1.5em}}
\label{fig:fun1}
\end{center}
\end{figure}

\begin{figure}[tbp]
\begin{center}
{\epsfig{file=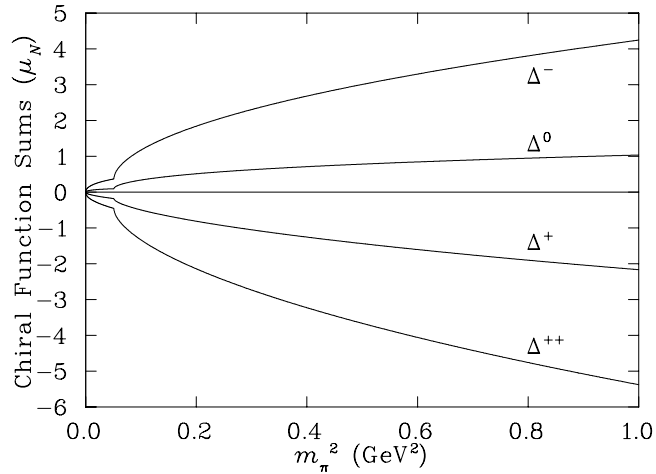, height=8.5cm, angle=90}}
\caption{Plots of the sum of all four chiral expansion terms of
Eq.~(\ref{PT2}), for each $\Delta$ baryon. 
\vspace{-1.5em}}
\label{fig:fun2}
\end{center}
\end{figure}

It is now recognized that in any extrapolation from the heavy quark
regime (where constituent quark properties are manifest) to the
physical world, it is imperative to incorporate the quark-mass
dependence of observables predicted by $\chi$PT in the chiral limit.
However, as results are often obtained using methods ideally suited for
heavy quark masses, it is imperative for the extrapolation function to
correctly reflect the behaviour of the physical observable in the
heavy quark mass regime as well.  An extrapolation function for the
$\Delta$-baryon magnetic moments satisfying these criteria is
\vspace{0.2em}
\begin{equation}
\mu = \frac{\mu_0}{1-{\Gamma(m_{\pi})}/{\mu_0}+\beta\ {m_{\pi}}^2}\, ,
\label{CEF1}
\end{equation}
where $\mu_0$ and $\beta$ are parameters optimized to fit results
obtained near the strange quark mass and $\Gamma(m_{\pi})$ is taken
from the chiral expansion for decuplet magnetic moments in
Sec.~\ref{sec:chiPT}
\begin{eqnarray}
\Gamma(m_\pi) &=& \chi_{\pi}\ \underbrace{m_{\pi}}_{A} + \chi_{K}\
                  \underbrace{(m_{K}-m_K^{(0)})}_{B}  
                  + \chi_{\pi}'\ \underbrace{
                  \left ( {\cal F}(-{\delta}_N,m_{\pi},{\mu_{\pi}}) 
                        - {\cal F}_{\pi} 
                  \right ) }_{C} \nonumber \\
               && + \chi_{K}'\ \underbrace{\left({\cal
                  F}(-{\delta}_N,m_{K},{\mu_{K}}) - {\cal F}_K
                  \right)}_{D}\, , 
\label{PT2}
\end{eqnarray}
where $m_K^{(0)}$, ${\cal F}_{\pi}$ and ${\cal F}_K$ are constants
defined to ensure that each term A, B, C and D, vanishes in the chiral
limit.  Utilizing the following relations provided by ${\chi}$PT
\begin{eqnarray}
{m_K}^2 &=& {m_K^{(0)}}^2 + \frac{1}{2} {m_{\pi}}^2 \, , \\
m_K^{(0)} &=& \sqrt{(m_K^{\mathrm{phys}})^2 -
\frac{1}{2}(m_{\pi}^{\mathrm{phys}})^2}\, , 
\end{eqnarray}
the four terms of Eq.~(\ref{PT2}) vanish in the chiral limit provided
\begin{eqnarray}
{\cal F}_{\pi} &=&   \frac{\delta}{\pi} \ln \left(\frac{(2
\delta)^2}{\mu^2}\right)\, , \nonumber \\ 
{\cal F}_K     &=& - \frac{\delta}{\pi} \ln
\left(\frac{(m_K^{(0)})^2}{\mu^2}\right) + \frac{2}{\pi}
\sqrt{(m_K^{(0)})^2-{\delta_N}^2} \left(\frac{\pi}{2} + \tan^{-1}
\frac{\delta_N}{\sqrt{(m_K^{(0)})^2 - {\delta_N}^2}}\right). 
\nonumber \\
\label{PT3}
\end{eqnarray}
Figure \ref{fig:fun1} presents a plot of each of the four terms of
Eq.~(\ref{PT2}) (without the chiral coefficient pre-factors) as a
function of $m_{\pi}^2$.  Figure \ref{fig:fun2} presents a plot of the
four terms summed with the appropriate weightings of
Table~\ref{table:DCC} for each of the four charge states of the
$\Delta$.

The extrapolation function of Eq.~(\ref{CEF1}) is designed to reproduce
the leading and next-to-leading nonanalytic structure expressed in
Eq.~(\ref{PT2}) for expansions about $m_{\pi}=0$.  
Eq.~(\ref{CEF1}) may be regarded as an analytic continuation of
Eq.~(\ref{PT2}), preserving the constraints imposed by chiral symmetry
and introducing the heavy quark mass regime behaviour to the
extrapolation function.  The LNA behaviour of Eq.~(\ref{PT2}) is
complemented by terms analytic in the quark mass with fit parameters
$\mu_0$ and $\beta$ adjusted to fit additional constraints on the
observable under investigation.

Hence the extrapolation function guarantees the correct nonanalytic
behaviour in the chiral limit.  Further as $m_{\pi}$ becomes large,
Eq.~(\ref{CEF1}) is proportional to $1/m_{\pi}^2$.  As ${m_{\pi}}^2
\propto m_q$ over the applicable mass range, the magnetic moment
extrapolation function decreases as 1/${m_q}$ for increasing quark
mass, precisely as the Dirac moment requires.  This extrapolation
function therefore provides a functional form bridging the heavy quark
mass regime and the chiral limit.

\section{Results}

The method employed to obtain our theoretical predictions is analogous
to that presented in our previous analysis of octet baryon magnetic
moments.\cite{CLT} We take the established input parameters, the
strange-constituent and strange-current quark masses ($M_s$ and
$c~m_s^{\rm phys}$ respectively\footnote{The parameter c is expected
to be the order of 1.}), obtained by optimizing agreement between the
AccessQM\footnote{The name indicates the mathematical
origins of the model: Analytic Continuation of the Chiral Expansion for
the SU(6) Simple Quark Model.}
and octet baryon magnetic moments.  There, $M_s = 565$ MeV
and $c~m_s^{\rm phys} = 144$ MeV provides optimal agreement.  

The constituent quark model (CQM) provides the following formulas that
relate the constituent quark masses to the delta magnetic moments.
\begin{equation}
\mu_{\Delta^{++}}=3 \mu_u\, , \quad \mu_{\Delta^{+}}=2 \mu_u +
\mu_d\, , \quad
\mu_{\Delta^{0}}= \mu_u + 2 \mu_d\, ,  \quad \mu_{\Delta^{-}}=
3\mu_d\, ,
\label{CQM}
\end{equation}
with
\begin{equation}
{\mu_u} = \frac{2}{3} \frac{M_N}{M_u}~{\mu_N}\, , \quad {\mu_d } = 
-\frac{1}{3} \frac{M_N}{M_d }~{\mu_N}\, , \quad
{\mu_s} = -\frac{1}{3} \frac{M_N}{M_s}~{\mu_N}\, .
\end{equation}
These formulas are used to obtain two magnetic moment data points near
the SU(3)-flavour limit where $u$, and $d$ quarks take values near the
$s$-quark mass.  

To fit Eq.~(\ref{CEF1}), which is a function of $m_{\pi}$, to the
magnetic moments given by the CQM in Eq.~(\ref{CQM}) with
constituent-quark masses $M_u=M_d=M_i~(i=1,2)$, we relate the pion
mass to the constituent quark mass via the current quark
mass.\cite{CLT}  Chiral symmetry provides
\be
\frac{m_q}{m^{\mathrm{phys}}_q}=\frac{m_{\pi}^2}
{(m_{\pi}^{\mathrm{phys}})^2}~~,
\label{equal}
\ee
where $m^{\mathrm{phys}}_q$ is the quark mass associated with the
physical pion mass, $m_{\pi}^{\mathrm{phys}}$.  From lattice studies,
we know that this relation holds well over a remarkably large regime
of pion masses, up to $m_{\pi} \sim 1$ GeV. The link between
constituent and current quark masses is provided by
\be
M = M_{\chi} + c~m_q\, ,
\ee
where $M_{\chi}$ is the constituent quark mass in the chiral limit and
$c$ is of order 1. Using Eq.~(\ref{equal}) this leads to
\be
M = M_{\chi} + 
\frac {c~m_q^{\mathrm{phys}}}{(m_{\pi}^{\mathrm{phys}})^2}\,
m_{\pi}^2\, .
\ee 
The link between the constituent quark masses $M_i$ and $m_{\pi}$ is
thus provided by
\be
m_{\pi \hspace{1.5pt} {i}}^2=(m_{\pi}^{\mathrm{phys}})^2~ \frac {M_{i}
-(M_{s}-c~m_{s}^{\mathrm{phys}})} {c~m^{\mathrm{phys}}_q} 
\hspace{15mm} (i = 1, 2)\, ,
\label{opt} 
\ee
where $M_{s}-{c~m^{\mathrm{phys}}_s}=M_{\chi}$ 
encapsulates information on the constituent 
quark mass in the chiral limit, and $c~m^{\mathrm{phys}}_s$ 
provides information on the strange current quark mass.
We use the ratio 
\be
\chi_{sq}= \frac {m_s^{\mathrm{phys}}}{m_q^{\mathrm{phys}}}=24.4 \pm
1.5\, ,
\ee
provided by $\chi$PT \cite{Leutwyler} to express the light current
quark mass, $m_q^{\mathrm{phys}}$, in terms of the strange current
quark mass, $m_s^{\mathrm{phys}}$, in Eq.~(\ref{opt}).

The analytic continuation of ${\chi}$PT, Eq.~(\ref{CEF1}), is fit to
the CQM as a function of $m_{\pi}^2$.  Results are presented in
Figs.~3 and 4.  The magnetic moments given by the CQM either side of
the SU(3)-flavour limit are indicated by a dot ($\bullet$) and the
theoretical prediction is indicated at the physical pion mass by a
star ($\star$).  These results, along with the parameters $\mu_0$ and
$\beta$ are summarized in Table~\ref{table:TEF}.

\begin{table}[tbp]
\tbl{Theoretical predictions for the charged $\Delta$ baryon magnetic
moments.  The fit parameters $\mu_0$ and $\beta$ are given for each
scenario.  The only known experimental value for the
$\Delta$ baryon magnetic moments is the $\Delta^{++}$ moment, 
recent measurements provide\protect\cite{Bosshard} $\mu_{\Delta^{++}}
= 4.52 \pm 50 \pm 45\ \mu_N$ and 
$\mu_{\Delta^{++}}= 6.14 \pm 51\ \mu_N$. \vspace*{1pt}} 
{\footnotesize
\tabcolsep24pt
\begin{tabular}{|l|c|c|c|}
\hline
{} &{} &{} &{} \\[-1.5ex]
Baryon         &$\mu_0$  &$\beta$    &AccessQM ($\mu_N$)  \\
\hline
{} &{} &{} &{} \\[-1.5ex]
$\Delta^{++}$  &+5.67     &0.16       &+5.39    \\
$\Delta^{+}$   &+2.69     &0.20       &+2.58    \\
$\Delta^{-}$   &$-$3.22   &0.06       &$-$2.99 \\[+1.5ex]
\hline
\end{tabular}
\label{table:TEF}
}
\end{table}

The interesting feature of these plots is the cusp at $m_{\pi}^2 =
\delta_N^2$ which indicates the opening of the octet decay channel,
$\Delta \to N \pi$.  The physics behind the cusp is intuitively
revealed by the relation between the derivative of the magnetic moment
with respect to $m_{\pi}^2$ and the derivative with respect to the
momentum transfer $q^2$, provided by the pion propagator
$1/(q^2+m_{\pi}^2)$ in the heavy baryon limit.  Derivatives with
respect to $q^2$ are proportional to the magnetic charge radius in the
limit $q^2 \to 0$,
\begin{equation}
\langle r_M^2\rangle \ =\ -6 \frac{\partial}{\partial q^2} G_M(q^2)
\arrowvert_{q^2=0}. 
\end{equation}  
If we consider for example $\Delta^{++} \to p\, \pi^+$ with $|j,
m_j\rangle =\ |3/2,\, 3/2\rangle$, the lowest-lying state conserving
parity and angular momentum will have a relative P-wave orbital
angular momentum with $|l,m_j\rangle \ =\ |1,1\rangle$.  Thus the
positively-charged pion makes a positive contribution to the magnetic
moment.  As the opening of the $p\, \pi^+$ decay channel is approached
from the heavy quark-mass regime, the range of the pion cloud
increases in accord with the Heisenberg uncertainty principle, $\Delta
E \, \Delta t \sim \hbar$.  Just above threshold the pion cloud extends
towards infinity as $\Delta E \to 0$ and the magnetic moment charge
radius diverges.  Similarly, $({\partial}/{\partial m_{\pi}^2}) G_M \to
-\infty$.  Below threshold, $G_M$ becomes complex and the magnetic
moment of the $\Delta$ is identified with the real part. The imaginary
part describes the physics associated with photon-pion coupling in
which the pion is subsequently observed as a decay product.

\begin{figure}[tbp]
\begin{center}
{\epsfig{file=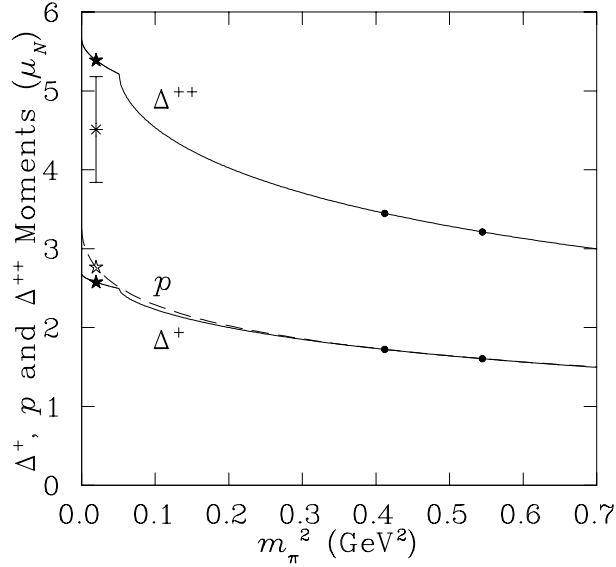, height=8cm, angle=90}}
\caption{The extrapolation function fit for $\Delta^{++}$ and
$\Delta^{+}$ magnetic moments.  The magnetic moments given by the CQM
either side of the SU(3)-flavour limit are indicated by dots
($\bullet$) and the theoretical prediction for each baryon is
indicated at the physical pion mass by a star ($\star$).  The only
available experimental data is for the $\Delta^{++}$ and is indicated
by an asterisk ($\ast$).  The proton extrapolation\protect\cite{CLT}
(dashed line) is included to illustrate the effect of the open decay
channel, $\Delta \to N\, \pi$, in the $\Delta^+$ extrapolation.  The
presence of this decay channel gives rise to a $\Delta^+$ moment
smaller than the proton moment. 
\vspace{-1em}}
\label{fig:d++}
\end{center}
\end{figure}

\begin{figure}[tbp]
\begin{center}
{\epsfig{file=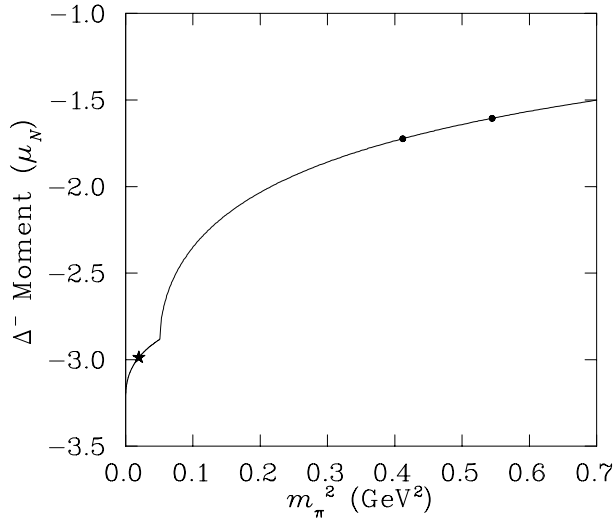, height=8cm, angle=90}}
\caption{The extrapolation function fit for the $\Delta^-$ magnetic
moment.  The magnetic moments given by the CQM either side of the
SU(3)-flavour limit are indicated by dots ($\bullet$) and the
theoretical prediction is indicated at the physical pion mass by a
star ($\star$).  There is currently no experimental value for the
$\Delta^-$ magnetic moment.
\vspace{-1em}}
\label{fig:d-}
\end{center}
\end{figure}

It is the NLNA terms of the chiral expansion for decuplet baryons that
contain the information regarding the decuplet to octet transitions.
These transitions are energetically favourable making them of
paramount importance in determining the physical properties of
$\Delta$ baryons.  The NLNA terms serve to enhance the magnitude of
the magnetic moment above the opening of the decay channel.  However,
as the decay channel opens and an imaginary part develops, the
magnitude of the real part of the magnetic moment is suppressed.  The
strength of the LNA terms, which enhance the magnetic moment magnitude
as the chiral limit is approach, overwhelms the NLNA contributions
such that the magnitude of the moments continues to rise towards the
chiral limit.

The inclusion of the NLNA structure into octet baryon magnetic moment
extrapolations is less important for two reasons.  The curvature
associated with the NLNA terms is negligible for the N and $\Sigma$
baryons and small for the $\Lambda$ and $\Xi$ baryons.  More
importantly one can infer the effects of the higher order terms of
$\chi$PT, usually dropped in truncating the chiral expansion, through
the consideration of phenomenological models.  If one incorporates
form factors at the meson-baryon vertices, reflecting the finite size
of the meson source, one finds that transitions from ground state
octet baryons to excited state baryons are suppressed relative to that
of $\chi$PT to finite order, where point-like couplings are taken.  In
$\chi$PT it is argued that the suppression of excited state
transitions comes about through higher order terms in the chiral
expansion.  As such, the inclusion of NLNA terms alone will result in
an overestimate of the transition contributions, unless one works very
near the chiral limit where higher order terms are indeed small. For
this reason octet to decuplet or higher excited state transitions have
been omitted in previous studies.\cite{CLT,LLT,HLT}

In the simplest CQM with $m_u = m_d$ the $\Delta^{+}$ and proton
moments are degenerate.  However, spin-dependent interactions between
constituent quarks will enhance the $\Delta^{+}$ relative to the
proton at large quark masses, and this is supported by lattice QCD
simulation results.\cite{LDW} As a result, early lattice QCD
predictions based on linear extrapolations\cite{LDW} report the
$\Delta^+$ moment to be greater than the proton moment.  However with
the extrapolations presented here which preserve the LNA behavior of
$\chi$PT, the opposite conclusion is reached.  We predict $\Delta^+$
and proton magnetic moments of 2.58~$\mu_N$ and 2.77~$\mu_N$
respectively.  The proton magnetic moment extrapolation\cite{CLT} is
included in Fig.~3 as an illustration of the importance of
incorporating the correct nonanalytic behaviour predicted by $\chi$PT
in any extrapolation to the physical world.  An experimentally
measured value for the $\Delta^{+}$ magnetic moment would offer
important insights into the role of spin-dependent forces and chiral
nonanalytic behaviour in the quark structure of baryon resonances.

\section{Conclusion}

An extrapolation function for the decuplet baryon magnetic moments has
been presented.  This function preserves the leading and
next-to-leading nonanalytic behaviour of chiral perturbation theory
while incorporating the Dirac-moment dependence for moderately heavy
quarks.  Interesting nonanalytic behavior in the magnetic moments
associated with the opening of the $\pi\, N$ decay channel has been
highlighted.  It will be interesting to apply these techniques to
existing and forthcoming lattice QCD results, and research in this
direction is currently in progress.

Experimental value exists only for the $\Delta^{++}$ magnetic
moment where the two most recent results are 
$\mu_{\Delta^{++}} = 4.52 \pm 0.50 \pm 0.45\ \mu_N$ and
$\mu_{\Delta^{++}} = 6.14 \pm 0.51\ \mu_N$. 
These
values are in good agreement with the prediction of 5.39~$\mu_N$ given
by our AccessQM as described above.  Arrival of experimental values
for the $\Delta^+$ and $\Delta^-$ magnetic moments are eagerly
anticipated and should be forthcoming in the next few years.  More
importantly, these techniques may be applied to the decuplet hyperon
resonances where the role of the kaon cloud becomes important.  We
look forward to new JHF results in this area in the future.

\vspace{-1em}
\section*{Acknowledgement}

This work was supported by the Australian Research Council.

\end{document}